\documentclass[11pt]{article}
\usepackage[utf8]{inputenc}
\usepackage{amsmath,amsfonts,amssymb,latexsym}
\usepackage[T1]{fontenc}
\newtheorem{satz}{Theorem}[section]

\newtheorem{defi}[satz]{Definition}

\newtheorem{bem}[satz]{Remark}
\newtheorem{lemma}[satz]{Lemma}
\newtheorem{koro}[satz]{Corollary}
\newtheorem{conclusion}[satz]{Conclusion}
\newtheorem{ob}[satz]{Observation}

\newtheorem{postulate}[satz]{Postulate}
\newtheorem{conjecture}[satz]{Conjecture}

\newcommand{\mcal}{\mathcal}

\newcommand{\tit}{\textit}

\newcommand{\beq}{\begin{equation}}
\newcommand{\eeq}{\end{equation}}
\newcommand{\RO}{\mcal{R}(\mcal{O})}
\newcommand{\M}{\mcal{M}}
\newcommand{\BH}{\mcal{B}(\mcal{H})}
\newcommand{\Ro}{\mcal{R}}
\newcommand{\rh}{\hat{\rho}_{\Omega}}
\newcommand{\HM}{\mcal{H}_{\M}}

\begin{document}
\thispagestyle{empty}
\begin{center}
\vspace*{1.0cm}
{\Large{\bf The Thermal Aspects of Relativistic Quantum Field Theory as an Observational Window in a Deeper Layer of Quantum Space-Time or: Dirac's Revenge}}
\vskip 1.5cm

{\large{\bf Manfred Requardt}}

\vskip 0.5cm

Institut fuer Theoretische Physik\\
Universitaet Goettingen\\
Friedrich-Hund-Platz 1\\
37077 Goettingen \quad Germany\\
(E-mail: requardt@theorie.physik.uni-goettingen.de)

\end{center}
 
\begin{abstract}
In this paper we shall derive the thermal properties of the relativistic quantum vacuum from a more primordial underlying structure which shares some properties with the old Dirac-sea picture. We show in particular how the Tomita-KMS structure in RQFT is a consequence of the structure and dynamics of the underlying pattern of vacuum fluctuations. We explain the origin of the doubling phenomenon in thermofield theory and the duality symmetry between a local algebra of observables and its commutant in RQFT and give an interpretation of the notion of thermal time. 

\end{abstract} \newpage
\setcounter{page}{1}
\section{Introduction}
It was observed some time ago (just to mention a few representative sources, see e.g. \cite{1},\cite{2},\cite{3},\cite{4},\cite{5}) that an accelerated observer, or put differently, an observer in the so-called \tit{Rindler wedge}, interprets the Minkowski vacuum as a \tit{thermal state}. The true nature of this phenomenon was and is still a topic of intense debate (cf., for example the seminal paper \cite{4} or the recent \cite{6}). We will comment on this issue in a forthcoming paper (\cite{7}).

Roughly at the same time this phenomenon was observed in another context (see \cite{8}). As a general source concerning this latter field we recommend 
\cite{9}. One of the advantages of this more general point of view is that it becomes obvious that such an (at first glance perhaps) surprising thermal behavior of states in the regime of relativistic quantum field theory (RQFT) is in fact a wide spread phenomenon, being the consequence of some rigorous results which hold true in RQFT. In all the phenomena under discussion a crucial role is played by the pure quantum phenomenon of \tit{entanglement} pervading the whole quantum world.

The physical basis is the celebrated \tit{Reeh-Schlieder theorem} (actually a case of strong entanglement of the \tit{quantum vacuum}); see e.g. \cite{9} or \cite{10}. (We note that for reasons of brevity we only cite generally available sources). It roughly says that one can generate the full Hilbert space of a RQFT by applying the fields or observables from a restricted  open domain of space-time (with non-empty causal complement) to the vacuum vector.

This observation allows us to apply a deep mathematical theorem from the field of v.Neumann operator algebras, i.e. the \tit{Tomita-Takesaki theorem}. The classic reference is \cite{11}. Readable accounts can also be found in \cite{9} and \cite{12}. Via this connection we are able to establish kind of a \tit{canonical dynamics} on v.Neumann algebras of observables, provided we are given an appropriate state. In RQFT the vacuum is such a state for any algebra of observables located in an open subset of space-time (with non-vanishing causal complement). This property was exploited in an important paper (\cite{13}) by Connes and Rovelli to conjecture some a priori thermal behavior and to introduce the concept of \tit{thermal time} in RQFT (see also, for example, \cite{14} and \cite{Martinetti}).

We dealt with related questions a little bit earlier in the paper \cite{15}, which circulated as a preprint but was never published (because we had the impression that certain points should be further clarified). A nice review of the whole field (including some comments to our earlier unpublished paper) is \cite{16}. The emphasis of our paper \cite{15} was not so much on rigorous mathematics but rather on the physical understanding of what was going on. In brief, we wanted to understand how an observer, being confined (together with his possible observations) to a subset of space-time, would describe the nature of the \tit{quantum vacuum} and the processes going on in this medium. This led us naturally to a scheme in which the quantum vacuum emerges as a thermal system of its own with important conceptual ingredient being the particular patterns of the \tit{vacuum fluctuations} (which we call \tit{pieces of the vacuum}; see below).

That the interpretation as a true thermal system is not entirely straightforward can be inferred from the following observation which we already advanced in \cite{15}. Take, for example, a finite system of quantum statistical mechanics in a \tit{non-equilibrium state}, described by a density matrix $\rho$. Under the ordinary time evolution  $\rho$ is hence not invariant. But writing  $\rho$ as
\begin{equation}\rho=e^{-K}    \end{equation}
which is always possible due to the positive definiteness of $\rho$ provided $\ln (\rho)$ is a well defined operator (which is usually the case modulo some technicalities), $\rho$ is now invariant under the new evolution $e^{isK}\circ e^{-isK}$, i.e.
\begin{equation}e^{isK}\cdot\rho\cdot e^{-isK}=\rho    \end{equation}
That is, under this new `time evolution', $\rho$ defines sort of an equilibrium state. More properly, it defines a so-called \tit{KMS-state} (KMS standing for Kubo-Martin-Schwinger, cf. e.g. \cite{9} or \cite{12}.

It can be proved that KMS-states share most of the properties of true equilibrium states. But it is sometimes overlooked that they do not automatically share a crucial and critical property. i.e. the property of \tit{return to equilibrium} (which depends in our view on the nature and distribution of eigenvalues of $\rho$ or $K$. Note that $K$ is in general \tit{not} a typical Hamilton operator). \tit{Return to equilibrium} is studied for example in \cite{17}. The whole topic has been discussed in some depth in \cite{16} (it was suggested by myself as supervisor).
\begin{bem}Such a property is unfortunately difficult to prove exactly in finite systems and in particular for systems describing RQFT in subsets of Minkowski space because of the \tit{quasi-periodicity} of correlation functions.  
\end{bem}

In the following we want to analyse the nature and origin of thermal behavior of such restricted systems in space-time in some detail with particular emphasis on the physical understanding. While we will use the important insights and results of algebraic quantum field theory (AQFT) as e.g. laid out in \cite{9}, we shall go beyond that and deal with the more primordial features of the quantum vacuum as a highly entangled and fluctuating medium. In this context we will develop and reveive the old \tit{Dirac-sea picture} of particle-hole excitations being excited in this medium. This is connected with the famous \tit{doubling} of modes in the \tit{thermo-field dynamics}. At the same time we will try to scrutinize the \tit{thermal time} picture mentioned above.

In a first step we shall construct almost isolated subsystems in RQFT together with their local dynamics (which is quite intricate due to the strong entanglement of the relativistic quantum vacuum). These systems have a certain similarity to the almost isolated systems of quantum statistical mechanics. We then distill the so-called \tit{pieces of the vacuum} which reflect the properties of the more primordial patterns of the vacuum fluctuations. In sect.5 we show how the \tit{Tomita-KMS-structure} emerges as a derived phenomenon from this more fundamental level. We show in particular how the \tit{doubling} of\tit{thermofield theory} and the \tit{duality-symmetry} between a local algebra of observables, $\RO$, and its commutant, $\RO'$, comes about by developing a picture which is reminiscent of the old \tit{Dirac-sea-picture}. In sect.6 we provide as a conclusion a coherent interpretative picture of our various observations.

\section{A Brief Introduction to Tomita-Takesaki Theory}
In this section we want to provide (for the convenience of the reader) a brief introduction to the so-called \tit{Tomita-Takesaki theory} of v.Neumann algebras. For the mathematical details see, for example, \cite{11},\cite{12},\cite{18}. As to the physical ramifications cf. \cite{9}. 

We deal with a v.Neumann algebra $\mcal{R}$ on a Hilbert space $\mcal{H}$ (e.g. the algebra of observables or fields, located in an open subset of space-time), having a \tit{cyclic} and \tit{separating} vector $\Omega$. Put differently, $\mcal{R}\Omega$ and $\mcal{R}'\Omega$ are both dense in $\mcal{H}$ with $\mcal{R}'$ the \tit{commutant algebra} of $\mcal{R}$, the latter property implying that $\Omega$ is separating for $\mcal{R}$, i.e.
\begin{equation} A\Omega=0\quad\text{for}\quad A\in \mcal{R}\quad\Rightarrow\quad A\equiv 0     \end{equation} 
In RQFT $\Omega$ does usually represent the vacuum vector.

The \tit{conjugate linear operator} $S$ ($S^2=1$) from $\mcal{R}\Omega$ to  $\mcal{R}\Omega$ 
\begin{equation} S\cdot A\Omega:=A^{\ast}\Omega     \end{equation}
is closable and has a \tit{polar decomposition}
\begin{equation}  S=J\cdot\Delta^{1/2}\quad ,\quad\Delta=S^{\ast}\cdot S\quad\text{positive}     \end{equation}
and $J$ \tit{antiunitary}, i.e.
\begin{equation} (J\psi|J\phi)=(\phi|\psi)   \end{equation}
hence
\begin{equation} J^{-1}=J^{\ast}=J\quad ,\quad J^2=1    \end{equation}
\begin{bem} For antilinear operators the adjoint is defined by
\begin{equation} (S^{\ast}\psi|\phi)=\overline{(\psi|S\phi)}=(S\phi|\psi)    \end{equation}
$S,\Delta$ are in the generic case unbounded.
\end{bem}
The same holds for the commutant $\mcal{R}'$ with $F$ instead of $S$.
\begin{equation} FA'\Omega=A'^{\ast}\Omega   \end{equation}

We have 
\beq F=S^{\ast}\quad\text{and}\quad F\cdot S=\Delta   \eeq
One can show that
\beq  \Delta^{-1/2}=J\Delta^{1/2}J  \eeq
(with $S=S^{-1}, F=F^{-1}$)
and
\beq  F=S^{\ast}=J\Delta^{-1/2}\quad ,\quad S\cdot F=\Delta^{-1}   \eeq
By definition ($S\Omega=S^{\ast}\Omega=\Omega$) we have
\beq  \Delta\Omega=\Omega=\Delta^{-1}\Omega\quad\Rightarrow J\Omega=\Omega   \eeq

If we write
\beq  \Delta=:e^{-K}   \eeq
and call $K$ the \tit{Tomita-Hamiltonian}, we can also define the one-dimensional group
\beq  U(s):=\Delta^{is}=e^{-isK}   \eeq
It follows from our previous results
\beq  J\Delta^{is}J=\Delta^{is}\quad ,\quad J\Delta^sJ=\Delta^{-s}   \eeq
The crucial properties of $\Delta^{is}$ and $J$ are 
\begin{satz}[Tomita] It holds
\beq  J\cdot\mcal{R}\cdot J=\mcal{R}'\; ,\; \Delta^{is}\cdot\mcal{R}\cdot\Delta^{-is}=\mcal{R}\; ,\; \Delta^{is}\cdot\mcal{R}'\cdot\Delta^{-is}=\mcal{R}'   \eeq 
\end{satz}
\begin{ob}[KMS-Property] With $\sigma_s:=\Delta^{is}\circ\Delta^{-is}$ the modular automorphism group it holds  for $A,B\in\mcal{R}$
\beq (\Omega|A_sB\Omega):=(\Omega|\sigma_s(A)B\Omega)=(\Omega|B\sigma_{s-i}(A)\Omega)=:(\Omega|BA_{s-i}\Omega)  \eeq
For $\mcal{R}'$ the modular group is
\beq  \sigma'_s:=\Delta^{-is}\circ\Delta^{is}=\sigma_{-s}  \eeq
\end{ob} 
Proof: we have
\beq (\Omega|\sigma_s(A)B\Omega)=(\Omega|AU^{\ast}(s)B\Omega)=(A^{\ast}\Omega|U^{\ast}(s)B\Omega)=(J\Delta^{1/2}A\Omega|U^{\ast}J\Delta^{1/2}B^{\ast}\Omega)  \eeq
which yields (with $J\Delta^{-is}=\Delta^{-is}J$):
\beq (J\Delta^{1/2}A\Omega|JU^{\ast}\Delta^{1/2}B^{\ast}\Omega) =(U^{\ast}\Delta^{1/2}B^{\ast}\Omega|\Delta^{1/2}A\Omega)=(\Omega|B\Delta^{1/2}U(s)\Delta^{1/2}A\Omega)  \eeq
and finally
\beq (\Omega|B\Delta^{i(s-i)}A\Omega)=(\Omega|B\sigma_{s-i}(A)\Omega)=:(\Omega|BA_{s-i}\Omega)   \eeq
\begin{bem} This is the generalization of the property, Gibbs-equilibrium states possess and is called the KMS-  (Kubo-Martin-Schwinger) property.
There is another sign convention in use which is motivated by physics, i.e. $\sigma_s(A):=\Delta^{-is}A\Delta^{is}$ with $\Delta^{-is}=e^{isK}$. This modifies also the KMS-condition which now reads
\beq  (\Omega|\sigma_s(A)B\Omega)=(\Omega|B\sigma_{s+i}(A)\Omega)   \eeq
\end{bem}
Note that in general the KMS-Hamiltonian is not an ordinary Hamiltonian as they occur in RQFT or many-body physics. While in the case of finite volume Gibbs states it is essentially a doubling of the usual Hamiltonian with two-sided spectrum, in the generic case it has an entirely different structure so that a property like \tit{return to equilibrium} need not hold. Anyway, the property of two-sidedness of the spectrum leads to interesting physical implications as e.g. a possible revival of the \tit{Dirac-sea picture}.
\begin{koro} From $\Delta\cdot\Omega=\Omega$ it follows
\beq (\Omega|\Delta^{is}A\Delta^{-is}\Omega)=(\Omega|A\Omega)   \eeq
\end{koro}
\section{Local, Almost Isolated Subsystems in RQFT, their Structure and Properties}
The central aims of our investigation are an analysis of the patterns and dynamics of the vacuum fluctuations in bounded open sets of space-time and their thermal properties. Second, as a consequence of this analysis, we want to scrutinize the \tit{thermal-time concept} of Connes and Rovelli, and, finally, we want to argue that as a result of all this there exists a, at first glance largely hidden, second (\tit{translocal}) structure beneath the surface structure with its causal behavior. We think that what makes quantum theory so different from ordinary classical and macroscopic physics comes mostly from this second, more primordial, structure. We have in particular in mind the puzzling phenomenon of \tit{entanglement}, which pervades allmost all of quantum physics. We will conjecture that it is related to the \tit{modular (Tomita) structure} of the local algebras of observables. 

There does however exist an obstacle which stands in the way of a straightforward analysis of the local aspects of (to a certainn degree, isolated) subsystems in RQFT, i.e. subsystems belonging to bounded open subsets, $\mcal{O}\in$ S-T (S-T denoting either space-time or Minkowski space in the following). We will denote the respective v.Neumann algebras by $\RO$. The problem is that, as a consequence of the unification of quantum theory and special relativity in RQFT, it is difficult to really isolate certain subsystems. This can be clearly seen from the content of the Reeh-Schlieder theorem, which holds for essentially every model of RQFT, implying that the fields and/or observables being located in open subsets of S-T (with non-empty causal complement) generate the full Hilbert space if being applied to the vacuum vector. The underlying reason is the existence of a strong (translocal) entanglement in the quantum vacuum. 
\begin{bem} The concept of locality can be easily formulated for fields and observables but is more tricky for states.
\end{bem}

As a consequence, almost all the constructions and approximations which work so well in, say, non-relativistic quantum theory to create (almost isolated) subsystems in an ambient medium (take e.g. the construction of the canonical ensemble of a subsystem from the microscopic ensemble of a larger system) do not work immediately in the context of RQFT or have to be dealt with with great care. A frequently applied tool is, for example, the \tit{tensoring construction} in the Unruh effect and related cases. One can of course establish a viable RQFT over a Hilbert space like $\mcal{H}_R\otimes\mcal{H}_L$ but this is (in general) not the restriction of the original RQFT in full Minkowski space to the right or left wedge. These are inequivalent representations and, by the same token, the Minkowski vacuum is not! a vector state in this tensor product. This can most easily be inferred from the following observation:
\begin{ob} The KMS- (or Tomita) evolution in the right or left wedge consists of Lorentz boosts, thus having a continuous spectrum. A vector in the tensor product yields density matrices in the left or right Hilbert spaces as reduced states. This leads to an evolution having a discrete spectrum and being related to the eigenvalues of the density matrices. We learn from this that  the full quantum vacuum is more complicated than a vector in the above tensor product.
\end{ob}
For more details see the following discussion, the general context can be found for example in \cite{Wald}.  See also the detailed analysis given in \cite{Kay1}.   

While such simplifying assumptions would make the analysis much simpler, it is in our view not even clear to what extent these are useful approximations. Therefore, in a first step, we will introduce some local structure in RQFT. Luckily, most of this has been already developed by AQFT and can be looked up in e.g. \cite{9}. To begin with, one usually wants to sharpen the so-called Einstein- or microcausality, i.e., that fields or observables (anti)commute if they are located in spacelike regions of S-T. On physical grounds one expects stronger causality properties, saying roughly that observations in spacelike regions are (statistically) independent. However, in general it holds even for spacelike regions that
\beq  (\Omega|AB\Omega) \neq (\Omega|A\Omega)\cdot (\Omega|B\Omega)\; ,\; A,B\;\text{spacelike}   \eeq
\begin{bem} In the following we assume that the local algebras, $\RO$ are factors, i.e. $\RO\cap\RO'=\lambda\cdot 1$. Furthermore, $\mcal{O}'$, the causal complement of $\mcal{O}$, is the interior of the set of points in S-T, lying spacelike to $\mcal{O}$.  
\end{bem}
A nice review about the various concepts of causal independence is \cite{Summers}.  

We choose the following concept, which was motivated in \cite{9}, sect. V\, 5.2. Given two algebras $\mcal{R}_1,\mcal{R}_2$, located in $\mcal{O}_1\subset\mcal{O}_2$ with $\mcal{O}'_1\cap\mcal{O}_2$ containing an open subset, hence $\mcal{R}_1\subset\mcal{R}_2$, there exists under natural assumptions a vector $\eta\in\mcal{H}$ which is cyclic and separating for $\mcal{R}_1\vee\mcal{R}'_2$ and which behaves as the vacuum, $\Omega$, on $\mcal{R}_1$ and $\mcal{R}_2'$. That is, it holds
\beq \label{Om} (\eta|AB'\eta)=(\Omega|A\Omega)\cdot (\Omega|B'\Omega)\;\text{for}\; A\in\mcal{R}_1,B'\in\mcal{R}'_2   \eeq
($\mcal{R}'_2$ the commutant of $\mcal{R}_2$, $\mcal{R}_1\vee\mcal{R}'_2$ the v.Neumann algebra, generated by  $\mcal{R}_1\cup\mcal{R}'_2$ ).
\begin{bem}Note that $\Omega$ is cyclic and separating for $\mcal{R}_1,\mcal{R}_2,\mcal{R}'_1\cap\mcal{R}_2$.
\end{bem}
The above property is equivalent with the two following equally important properties.
\begin{satz} The above property and the two following properties are equivalent.\\
i) There exists a unitary operator $W:\mcal{H}\rightarrow \mcal{H}\otimes\mcal{H}$ with
\beq WAB'W^{\ast}=A\otimes B'\;\text{and}\; WAB'\eta=A\Omega\otimes B'\Omega   \eeq
in particular
\beq W\eta=\Omega\otimes\Omega  \eeq
ii) There exists a v.Neumann factor $\mcal{M}$ between $\mcal{R}_1$ and $\mcal{R}_2$, i.e.
\beq  \mcal{R}_2\supset\mcal{M}\supset\mcal{R}_1  \eeq
which is of type I.
\end{satz}
\begin{bem} The unitarity of $W$ follows from the fact that $\eta,\Omega$ are both cyclic and separating for $\mcal{R}_1\cup\mcal{R}_2'$ and formula (\ref{Om}).
\end{bem} 
We are not going to explain the \tit{type-theory} of operator algebras (cf. the above cited literature on operator algebras). Suffice it to say that our $\mcal{M}$ is of type $I^{\infty}$, which means the following:
\begin{satz} $\mcal{M}$ is unitarily equivalent to a $\mcal{B}(\mcal{H})$, more specifically:
\beq \mcal{M}=W^{\ast}(\mcal{B}(\mcal{H}\otimes 1)W\quad \mcal{M}'=W^{\ast}(1\otimes\mcal{B}(\mcal{H})W   \eeq
\end{satz}
For a proof see e.g. \cite{9},\cite{Dixmier} p.124 or \cite{Jones}.
\begin{bem} The structure of the algebras $\RO$ is usually much more complicated and we will restrict ourselves in the following primarily to the physical analysis of the algebra $\mcal{M}$ which is more akin to the type of system we are accustomed to in ordinary quantum statistical mechanics.
\end{bem}
\begin{ob} It will turn out that $\M$ describes an almost isolated subsystem in Minkowski space, being defined by a set of boundary conditions in very much the way we create isolated subsystems in quantum statistical mechanics via, for example, boundary conditions being imposed on some subsystem Hamiltonian.
\end{ob}   

Evidently it holds
\beq \mcal{M}\subset \mcal{R}_2\;\text{and}\; \mcal{M}'\supset \mcal{R}'_2   \eeq
In the following we study the properties of $\mcal{M}$ and  $\mcal{M}'$. As both are \tit{factors} as unitary images of $\mcal{B}(\mcal{H})\otimes 1$ and $1\otimes\mcal{B}(\mcal{H})$, it holds 
\beq \mcal{M}\cap\mcal{M}'=\lambda\cdot 1    \eeq
and therefore 
\begin{lemma} $\mcal{M}\cup\mcal{M'}$ generate an irreducible algebra in $\mcal{B}(\mcal{H})$. As the commutant is $\lambda\cdot 1$ the weak closure $\mcal{M}\vee\mcal{M}'$ is the full  $\mcal{B}(\mcal{H})$ (the famous v.Neumann bicommutant theorem).
\end{lemma}
\begin{ob} This implies physically that the full observable information in S-T is encoded in  $\mcal{M}\vee\mcal{M}'$.
\end{ob}

It is of tantamount importance to understand in what sense $\mcal{M}$ contains more elements than $\mcal{R}_1$ and what are the characteristic properties of the above mentioned \tit{boundary conditions}. In the seminal paper \cite{Longo} this is discussed in a more general context on p.511 and as a reference \cite{11} is mentioned. As we had difficulties to find the corresponding information in \cite{11} (it is possibly hidden in a more general and abstract result) we will provide our own proof below. We begin with some preparatory steps which will be of general relevance for our further analysis.  We have
\beq W\mcal{R}_1W^{\ast}=\mcal{R}_1\otimes 1  \eeq
In the same way $\Omega$ is mapped on a vector $\hat{\Omega}$ in $\mcal{H}\otimes\mcal{H}$. With $\{e_j'\}$ an ON-basis in $\mcal{H}$ we have
\beq \hat{\Omega}:=W\Omega=\sum_{ij} c_{ij}e_i'\otimes e_j'\in\mcal{H}\otimes\mcal{H}   \eeq
With $A\in\mcal{M}$ and $\hat{A}:=WAW^{\ast}\in\BH\otimes 1$ we have
\beq (\Omega|A\Omega)=(\hat{\Omega}|\hat{A}\hat{\Omega})  \eeq

In the usual way we can reduce the pure vector state $\hat{\Omega}\in \mcal{H}\otimes\mcal{H}$ to a density matrix over $\BH\otimes 1$.
\begin{ob} We have for $A\in\M$
\beq (\Omega|A\Omega)= (\hat{\Omega}|\hat{A}\hat{\Omega})=Tr(\rho\hat{A})   \eeq
\end{ob}
where, by abuse of notation, we identify the element $\hat{A}\in\BH\otimes 1$ with the corresponding element in the first factor. $\rho$ is a (positive definite) \tit{density matrix} in $\BH$ with matrix elements (in the basis $\{e_i'\}$)
\beq b_{ij}=\sum \overline{c}_{il}c_{jl}  \eeq
\begin{bem} The positive definiteness follows from the positive scalar product $(\hat{\Omega}|\circ\hat{\Omega})$.
\end{bem}

It is useful and common practice to choose a so-called \tit{Schmidt-basis} $\{e_i\otimes f_i\}$ in $\mcal{H}\otimes\mcal{H}$ instead of the general basis $\{e_i'\otimes e_j'\}$. This choice diagonalizes the density matrix $\rho$.
\begin{ob} In the Schmidt-basis we have
\beq \hat{\Omega}=\sum_j e^{-\lambda_j/2}e_j\otimes f_j\;\text{and}\;\rho=\sum_j e^{-\lambda_j}|e_j><e_j|  \eeq
(As the $w_j$ are positive with $w_j\leq 1$, it exists a representation $w_j= e^{-\lambda_j/2}$ with $\lambda_j\geq 0$).
Hence
\beq (\Omega|A\Omega)=\sum_j e^{-\lambda_j}(e_j|\hat{A}e_j)\;\text{for}\;A\in\M  \eeq
By the same token there exists a density matrix for $\M'$, i.e.
\beq (\Omega|B\Omega)=Tr(\tilde{\rho}\hat{B})=\sum_j e^{-\lambda_j}(f_j|\hat{A}f_j)\;\text{for}\;B\in\M'  \eeq
\end{ob}
\begin{bem} We show below that $\rho$ is invertible, i.e. all the eigenvalues are non-vanishing. Furthermore, it is a consequence of the Schmidt representation that $\rho$ and $\tilde{\rho}$ have the same eigenvalues with the same multiplicities (apart from possible null-eigenspaces).
\end{bem}
Proofs of this useful representation can be found in many places. A classical source is \cite{Neumann}. A modern concise version can be found in e.g. \cite{Requ}.

We now employ that the vacuum $\Omega$ is cyclic and separating for\\
 $\Ro_1,\Ro_2,\Ro_1',\Ro_2'$ with $\Ro_2\supset\Ro_1$. 
\begin{ob} $\Omega$ is cyclic and separating for $\M$ and $\M'$. 
\end{ob}
From this follows immediately
\begin{ob} $\hat{\Omega}$ is cyclic and separating for $\BH\otimes 1$ and $1\otimes\BH$.
\end{ob}
\begin{koro} This implies that $\rho$ and $\tilde{\rho}$ are invertible operators.
\end{koro}
(Such properties were for example discussed and exploited in \cite{16}, sect. 7.13).
Now comes an important observation:
\begin{ob} $\rho\otimes\tilde{\rho}^{-1}$ leaves the vector $\hat{\Omega}$ invariant, i.e.:
\beq \rho\otimes\tilde{\rho}^{-1}\cdot\hat{\Omega}=\hat{\Omega}   \eeq
\end{ob}
Proof: Use the representation of  $\hat{\Omega}$ in the Schmidt-basis and the corresponding spectral representation of   $\rho\otimes\tilde{\rho}^{-1}$.

We now come back to the structure of $\M$.
\begin{satz} $\M$ is generated by $\Ro_1$ and the observable 
\beq \hat{\rho}:=W^{\ast}(\rho\otimes 1)W\in \M   \eeq
\end{satz}
Proof: In a first step we show that $\Ro_1$ and $\rho$ generate an irreducible algebra in $\BH$. We assume there exists an element $B$ which commutes with $\rho$ and $\Ro_1$. Then $B\otimes 1$ trivially commutes with $\rho\otimes\tilde{\rho}^{-1}$. From general spectral theory it follows that it commutes with the spectral projectors of  $\rho\otimes\tilde{\rho}^{-1}$, i.e.:
\begin{lemma} Such a  $B\otimes 1$ commutes with $P_{\hat{\Omega}}$.
\end{lemma}
which yields
\beq (B\otimes 1)\hat{\Omega}= (B\otimes 1)P_{\hat{\Omega}}\hat{\Omega}=P_{\hat{\Omega}} (B\otimes 1)\hat{\Omega}=c\cdot\hat{\Omega}  \eeq

Furthermore, $B\otimes 1$ is assumed to commute with $\Ro_1\otimes 1$. But $\hat{\Omega}$ is cyclic for $\Ro_1\otimes 1$, hence
\beq (B\otimes 1)\cdot (\Ro_1\otimes 1)\hat{\Omega}=(\Ro_1\otimes 1) (B\otimes 1)\hat{\Omega}=c\cdot  (\Ro_1\otimes 1)\hat{\Omega}  \eeq
\begin{lemma} we have 
\beq  (B\otimes 1)=c\cdot 1\otimes 1\;\text{and}\; B=c\cdot 1  \eeq
\end{lemma}
Hence we have finally
\begin{satz} $\Ro_1\cup\rho$ generate an irreducible algebra in $\BH$. Thus, its weak closure is the full  $\BH$.
Correspondingly, $\BH\otimes 1$ is the weak closure of $( \Ro_1\otimes 1)\cup (\rho\otimes 1)$.
\end{satz}
Transferring this result back to $\M$ with the help of $W^{\ast}\circ W$, we see that $\M$ is generated by $\Ro_1$ and $\hat{\rho}$ which proves the above theorem.
\begin{koro} In the same way, $\BH\otimes 1$ and $\M$ are generated by $(\mcal{R}_1\otimes 1),\mcal{R}_1$ and the spectral projections of $\rho\otimes 1$ and $\hat{\rho}$, i.e. $\{W^{\ast}(P_{e_i}\otimes 1)W\}$. Note that $\mcal{R}_1$ is mapped under $W\circ W^{\ast}$ onto $\mcal{R}_1\otimes 1$.
\end{koro}
\begin{bem} One should note that a similar result holds in \tit{Axiomatic Quantum Field Theory}. It can be proved that a local algebra $\RO$ together with the projection on the vacuum generates an irreducible algebra (see \cite{10}). Furthermore, $\rho$ is not! a local element but $\hat{\rho}$ is contained in $\M\subset\Ro_2$.
\end{bem}
\section{The Thermal Properties of Localized Subsystem and the ``Pieces of the Vacuum''}
It is now straightforward to introduce a dynamics on $\M$ and $\M'$ which turns out to be the modular (Tomita) evolution. We show that
\begin{ob} \label{J} It holds
\beq \Delta=\rho\otimes\tilde{\rho}^{-1}\; ,\; \Delta\cdot\hat{\Omega}=\hat{\Omega}\; ,\;\Delta(e_i\otimes f_j)=e^{-(\lambda_i-\lambda_j)}(e_i\otimes f_j)   \eeq
is the modular operator on $\mcal{H}\otimes\mcal{H}$ with
\beq \Delta^{is}=\rho^{is}\otimes\tilde{\rho}^{-is}    \eeq
and modular Hamiltonian
\beq K:=-\ln (\rho\otimes\tilde{\rho}^{-1})=K_1\otimes 1-1\otimes K_2\; ,\; K(e_i\otimes f_j)=(\lambda_i-\lambda_j)(e_i\otimes f_j)     \eeq
with 
\beq \rho=e^{-K_1}\; ,\; \tilde{\rho}=e^{-K_2}   \eeq
The unitary evolution $ \Delta^{is}\circ\Delta^{-is}$ leaves $\BH\otimes 1$ and $1\otimes\BH$ separately invariant. The antiunitary conjugation $J$ is given by
\beq J(e_i\otimes f_j):=e_j\otimes f_i  \eeq
and antilinear continuation on arbitrary vectors.
\end{ob}
Proof: It is sufficient to show that $\Delta^{is}\circ\Delta^{-is}$ fulfills the KMS-condition on $\BH\otimes 1$  with $\Delta^{-is}\circ\Delta^{is}$ fulfilling the KMS-condition on $1\otimes\BH$. That is, with $A,B\in\BH\otimes 1$:
\beq (\hat{\Omega}|\sigma_s(A)B\hat{\Omega})= (\hat{\Omega}|B\sigma_{(s-i)}(A)\hat{\Omega})   \eeq
This is an easy exercise (cf. e.g. \cite{16}sect.7.13). To show that $J$ is the Tomita conjugation we show that $J\Delta^{1/2}A\Omega=A^{\ast}\Omega$ in the basis $\{e_i\otimes f_j\}$ using the explicit form of $\Omega$ and the representation of $\Delta$ in this basis. Note that in a basis $A\rightarrow A_{ij}\Rightarrow A^{\ast}\rightarrow \overline{A}_{ji}$.
\begin{bem} For technical reasons ($\Delta$ is unbounded) $A$ has to fulfill a technical condition; see e.g. \cite{Bratteli}.
\end{bem}

In the same way we can transport the KMS-property to $\M$ and $\M'$ with the help of $W^{\ast}\circ W$.
\begin{ob} It follows that $\hat{\rho}\cdot\hat{\tilde{\rho}}^{-1}, \hat{\rho}^{-1}\cdot\hat{\tilde{\rho}}$ define the Tomita (modular) operators on  $\M$ and $\M'$ with
\beq \hat{\rho}\cdot\hat{\tilde{\rho}}^{-1}:=W^{\ast}(\rho\otimes 1)W\cdot W^{\ast}(1\otimes\tilde{\rho}^{-1})W   \eeq
and $\hat{\rho},\hat{\tilde{\rho}}$ living in $\M$ and $\M'$.
\end{ob}
The corresponding expectation values with observables from $\M$ or $\M'$ fulfill the KMS-condition relative to $\Omega$.

We come now to the localization properties of suitable states. We will construct a localized subspace in $\mcal{H}$ and what we call the \tit{pieces of the vacuum} (the existence of which we already motivated heuristically in \cite{15} without having the complete technical machinery at out disposal). We define the subspace
\beq \mcal{H}_{\M}:= W^{\ast}(\mcal{H}\otimes\Omega)\subset\mcal{H}   \eeq
The projector on $\mcal{H}_{\M}$ is
\beq P_{\M}=W^{\ast}(1\otimes P_{\Omega})W\in\M'   \eeq
We have (by construction, see the beginning of sect.3):
\beq \eta=W^{\ast}(\Omega\otimes\Omega)  \eeq
hence
 \beq W^{\ast}(\mcal{H}\otimes\Omega)=W^{\ast}(\BH\otimes 1)WW^{\ast}(\Omega\otimes\Omega)=\M\cdot\eta  \eeq
It therefore holds
\begin{ob} We have
\beq \mcal{H}_{\M}:=W^{\ast}(\mcal{H}\otimes\Omega)=\M\cdot\eta  \eeq
(cf. sect. V5.3 in \cite{9}).
\end{ob}

Localized states in RQFT are defined in the following way (see \cite{9} and further references given there):
\begin{defi} A state $\psi$ is localized in $\mcal{O}\subset S-T$ if for any observable $B'\in\RO'$ it holds:
\beq (\psi|B'\psi)=(\Omega|B'\Omega)  \eeq
that is, $\psi$ looks like the vacuum for space-like observations relative to $\mcal{O}$.
\end{defi}
We take now a state $W^{\ast}(\psi\otimes\Omega)$, $\psi$ a unit vector, from the subspace $\M\eta=W^{\ast}(\mcal{H}\otimes\Omega)$ and calculate with $B_2'\in\mcal{R}'_2$:
\begin{multline} (W^{\ast}(\psi\otimes\Omega)|B_2'\cdot W^{\ast}(\psi\otimes\Omega)'W^{\ast}(\psi\otimes\Omega)=((\psi\otimes\Omega)|WB_2'W^{\ast}(\psi\otimes\Omega))=\\  ((\psi\otimes\Omega)|(1\otimes B_2'(\psi\otimes\Omega))=(\Omega|B_2'\Omega)  \end{multline}
(exploiting the property of the map $W\circ W^{\ast}$; note that $B_2'\in\M'$).
\begin{ob} States in $\mcal{H}_{\M}$ are localized in $\mcal{O}_2\supset\mcal{O}_1$.
\end{ob}
\begin{bem} Note that $\M\Omega$ still spans the full Hilbert space $\mcal{H}$ as $\Omega$ is cyclic for $\M$. While $\eta$ is cyclic for $\mcal{R}_2'\vee\mcal{R}_1$, it is, in contrast to $\Omega$, not cyclic for $\mcal{R}_2'$ or $\mcal{R}_1$ separately.
\end{bem} 

We now will construct particular localized states which we will call \tit{pieces of the vacuum}. We take elements $e_i\in\mcal{H}$ of the above Schmidt-basis of $\rho$. For each $e_i$ we can find a (non-unique) unitary operator $U_i\in\BH$ with 
\beq  U_i\cdot\Omega=e_i   \eeq
We get
\beq W^{\ast}(e_i\otimes\Omega)=W^{\ast}((U_i\otimes 1)(\Omega\otimes\Omega))=W^{\ast}(U_i\otimes 1)W\cdot W^{\ast}(\Omega\otimes\Omega)=\hat{U_i}\cdot\eta  \eeq
with $\hat{U_i}$ a unitary operator in $\M$. Evidently, $\hat{U_i}\cdot\eta$ is a state localized in $\mcal{O}_2$.

We have with $A\in\M\; ,\;\hat{A}=WAW^{\ast}\in\BH\otimes 1$:
\begin{multline} (\Omega|A\Omega)=Tr(\rho\hat{A})=\sum_j e^{-\lambda_j}(e_j|\hat{A}e_j)= \sum_j e^{-\lambda_j}((e_j\otimes\Omega)|\hat{A}(e_j\otimes\Omega))=\\ \sum_j e^{-\lambda_j}(W^{\ast}(e_j\otimes\Omega)|AW^{\ast}(e_j\otimes\Omega))= \sum_j e^{-\lambda_j}(\hat{U}_j\eta|A\hat{U}_j\eta)=\\ \sum_j e^{-\lambda_j}(\psi_j|A\psi_j)   \end{multline}
with $\psi_j:=\hat{U}_j\eta\in \mcal{H}_{\M}$ localized in $\mcal{O}_2$.
\begin{ob} The above calculation shows that the localized states $\{\psi_j\}$ can be regarded as pieces of the vacuum.
\end{ob}
\begin{lemma} The $\{\psi_j\}$ span $\mcal{H}_{\M}$ as an ON-basis. Furthermore,  $\mcal{H}_{\M}$ is a true subspace of $\mcal{H}$.
\end{lemma}
Proof: $\{e_j\otimes\Omega\}$ span $\mcal{H}\otimes\Omega$, hence  $\{W^{\ast}(e_j\otimes\Omega)\}$ span  $\mcal{H}_{\M}$. $\mcal{H}\otimes\Omega$ is orthogonal to $\mcal{H}\otimes\phi$ with $\phi\perp\Omega$. Hence
\beq W^{\ast}(\mcal{H}\otimes\Omega)\perp W^{\ast}(\mcal{H}\otimes\phi)   \eeq

We showed above that $W^{\ast}(\rho^{is}\otimes\tilde{\rho}^{-is})W$ is the Tomita and KMS-evolution relative to $(\M,\M',\Omega)$. On the other hand, $W^{\ast}(\rho\otimes 1)W$ defines the same evolution on $\M$ without leaving however the vacuum $\Omega$ invariant, i.e.:
\beq \alpha_s(A)=\hat{\rho}^{is}A\hat{\rho}^{-is}\;\text{with}\;\hat{\rho}:=W^{\ast}(\rho\otimes 1)W \eeq
(see above).

We now undertake to formulate statistical thermodynamics together with a (thermal) evolution on the localized subsystem $\M$ on $\HM$ as it is commonly formulated for finite volume Gibbs ensembles. Note that $\rho\otimes 1,W^{\ast}(\rho\otimes 1)W$ are not! density matrices as all eigenspaces are obviously infinitely degenerated. On the other hand, $\rho\otimes P_{\Omega}$ is a density matrix with only the null-eigenspace being infinitely degenerated. The same holds for
\beq  \rh:=W^{\ast}(\rho\otimes P_{\Omega})W   \eeq
which, however, is no longer! an element of $\M$.
\begin{ob} It holds
\beq  \rho\otimes P_{\Omega}\!\upharpoonright_{\mcal{H}\otimes\Omega}\;\text{and}\;\rh\!\upharpoonright_{\HM}   \eeq
are invertible density matrices if we restrict them to the subspaces
\beq \mcal{H}\otimes\Omega\;\text{and}\; W^{\ast}(\mcal{H}\otimes\Omega)=\HM   \eeq
The eigenvectors of $\rh\!\upharpoonright_{\HM}$ are $\{\psi_j\}$ with eigenvalues $\{e^{-\lambda_j}\}$. We have for $A\in\M$:
\beq  (\Omega|A\Omega)=Tr(\rh\cdot A)=\sum_j e^{-\lambda_j}\cdot (\psi_j|A\psi_j)   \eeq
Evidently, $\M$ leaves $\HM=\M\cdot\eta$ invariant.
\end{ob}

We define the corresponding evolution
\beq W^{\ast}(\rho^{is}\otimes P_{\Omega}^{is})W=\rh^{is}   \eeq
\begin{ob} $\rh^{is}$ leaves $\HM$ invariant:
\beq (\rho^{is}\otimes P_{\Omega}^{is})\cdot (\psi\otimes\Omega)=\rho^{is}\psi\otimes e^{is}\Omega=e^{is}\cdot\rho^{is}\psi\otimes\Omega  \eeq
and correspondingly
\beq  W^{\ast}(\rho^{is}\otimes P_{\Omega}^{is})WW^{\ast}(\psi\otimes\Omega)=W^{\ast}e^{is}\cdot\rho^{is}\psi\otimes\Omega)\in\HM  \eeq
In the same way, for $A\in\M$:
\beq \rh^{is}\cdot A\cdot\rh^{-is}=W^{\ast}(\rho^{is}\hat{A}\rho^{-is}\otimes 1)W\in\M  \eeq
\end{ob}

Finally, as $\rh\!\upharpoonright_{\HM}$ is invertible, we can prove the KMS-property:
\begin{ob} For $A,B\in\M$ we have
\beq Tr(\rh\rh^{is}A\rh^{-is}B)=Tr(\rh^{-1}\rh B\rh^{i(s-i)}A\rh^{-is})=Tr(\rh B\rh^{i(s-i)}A\rh^{-i(s-i)})  \eeq
\end{ob}
\begin{conclusion} We have shown that $(\rh\upharpoonright\!_{\HM},\M,\rh^{is}\circ\rh^{-is})$ behaves exactly in the same way as a finite thermodynamical system as we are accustomed to in ordinary quantum statistical mechanics. This proves what we already conjectured in \cite{15}.
\end{conclusion} 
Analogously we can define
\beq \mcal{H}_{\mcal{M}'}:=W^{\ast}(\Omega\otimes\mcal{H})    \eeq
$ \mcal{H}_{\mcal{M}'}$ is localized in $\mcal{O}_1'$. With 
\beq  U_j'\Omega=f_j\; ,\;\psi_j':=\hat{U}_j'\eta=W^{\ast}(\Omega\otimes f_j)  \eeq
we have for $B'\in\M'$
\beq  (\Omega|B'\Omega)=Tr(\hat{\tilde{\rho}}_{\Omega}\cdot B')=\sum_j e^{-\lambda_j}\cdot (\psi_j'|B'\psi_j')   \eeq
with
\beq  \hat{\tilde{\rho}}_{\Omega}:=W^{\ast}(P_{\Omega}\otimes\tilde{\rho})W   \eeq
\begin{ob} We can now define the Tomita evolution on $\M\otimes\M'$ and $\HM\otimes\mcal{H}_{\mcal{M}'}$. It reads:
\beq \hat{\Delta}_{\Omega}:=e^{-\hat{K}_{\Omega}}:=      \rh\otimes\hat{\tilde{\rho}}_{\Omega}^{-1}\;\text{for}\;\M  \eeq
Its eigenvectors are $(\psi_i\otimes\psi_j')$ with eigenvalues $e^{-(\lambda_i-\lambda_j)}$.
The Tomita operator for $\M'$ is $ \hat{\Delta}_{\Omega}^{-1}$. The Tomita conjugation $J$ is given by
\beq J(\psi_i\otimes\psi_j'):=\psi_j\otimes\psi_i'  \eeq
and antilinear continuation on general vectors (for the proof cf. observation \ref{J}). 
\end{ob}
\begin{bem} We want to mention in this context the paper \cite{Kay2}, in which so-called double-KMS-states are introduced having a similar tensor product structure (see also \cite{Kay1}).
\end{bem}

\section{A Physical ``Derivation'' of the Tomita-Takesaki Theory and the Thermal Properties of the Relativistic Quantum Vacuum via a Generalized Dirac-Sea Picture}
In ordinary RQFT the vacuum is typically introduced as the lowest (Poincare invariant) energy state in the Hilbert space with some of its properties then (somewhat indirectly) rederived via correlations of fields and observables. There does exist however another less developed philosophy which says that the quantum vacuum is actually the primordial substratum from which everything else ultimately emerges (a point of view which was for example promoted by Wheeler, see \cite{Wheeler}. This is also our point of view and research program (see, for example, \cite{Wormhole},\cite{Quantum},\cite{Planck} and references therein). What is, to mention one particular point, remarkable is its strong and long-range \tit{quantum entanglement} which is, furthermore, of a certain \tit{translocal} character (see \cite{Ado1},\cite{Ado2} and our above mentioned papers). 

In the following we want to argue that the dynamics and the patterns of the vacuum fluctuations are responsible for the surprising thermal properties of the vacuum state. In a first step we want to show that beside the ordinary Minkowski-space representation of states and observables there exists a dual but, in our view, more primordial description of the processes, going on in RQFT and in particular in the quantum vacuum and that the Tomita-Takesaki theory is the mathematical implementation of this more hidden structure on the level of Minkowski-space.

It is well known that the phenomenon of \tit{doubling} of the quantum field structure in \tit{thermal field theory} is at first glance a formal effect of the description of thermal effects within a Hilbert-space framework (cf. \cite{Thermal}). Schroer called it (in the context of AQFT) the \tit{shadow algebra} (\cite{Schroer}, footnote on p.5 and p.7). To our knowledge such a doubling structure was observed for the first time in the statistical mechanics of the free Bose- and Fermi-gas in the infinite volume limit by Araki et al. (\cite{Araki1} and \cite{Araki2}). The connection between thermofield dynamics and the Tomita-KMS theory was already analyzed by Ojima in \cite{Ojima}. We have seen that a similar structure does prevail in the Tomita-Takesaki theory when applied to the v.Neumann algebra of observables $\RO$ of a finite region in Minkowski-space. We observe a related but slightly different situation in the analysis of the Unruh-effect or the \tit{interior-exterior-relation} in black hole physics. In all these cases we find a doubling structure being accompanied by thermal behavior. 

In the Tomita-Takesaki theory the two most remarkable phenomena are, first, the apriori existence of an automorphism group $\sigma_s= \Delta^{is}\circ\Delta^{-is}$ of the algebra $\RO$,which obeys the KMS-condition, which, on its side, is the generalization of the Gibbs-equilibrium condition of a box-system and Hamiltonian. And secondly, the apriori (antiunitary) symmetry between $\RO$ and its commutant (causal complement) $\RO'$ induced by the operator $J$. The first property can be related to the \tit{thermal time concept} of Connes-Rovelli. The second property is related to the doubling phenomenon. In the following we undertake to explain the occurrence of these two phenomena and derive them from a more primordial description of the relativistic quantum vacuum. Mathematically $\Delta$ and $J$ are related via
\beq S=J\Delta^{1/2}\; ,\; F=J\Delta^{-1/2}   \eeq
with
\beq SA\Omega=A^{\ast}\Omega\; ,\; FB'\Omega=(B')^{\ast}\Omega   \eeq
($A\in\RO\; ,\; B'\in\RO'$). We show that they are also physically related.

What makes a realistic interpretation of the results of the Tomita-Takesaki theory so complicated is the following observation. For a canonical (box) multiparticle system in a heat bath (i.e., the canonical Gibbs ensemble) we can provide an interpretation of both the automorphism group (it is essentially the ordinary Hamiltonian time evolution) and the doubling phenomenon (which is somewhat less obvious; see below). On the other hand, for a v.Neumann algebra of observables $\RO$, being embedded in infinite Minkowski space and with the vacuum $\Omega$ playing the role of the Gibbs-equilibrium state, the physical nature of the Tomita-evolution is (apart from a few cases like the \tit{Rindler-wedge}) obscure. That holds even more so for the symmetry between $\RO$ and its commutant $\RO'$ which, in contrast to the box-system, live in geometrically well-separated regions of S-T.
\subsection{Doubling in the Box-Canonical Ensemble}
In the box-canonical ensemble the physical explanation of doubling goes as follows. While in the ordinary statistical mechanics literature the equilibrium state is traditionally represented as a density matrix one has to switch the point of view a little bit. That is, as in the framework, presented in e.g. \cite{9} or in \cite{Thermal}, one represents the equilibrium state as a Hilbert space vector. Then, in case one is familiar with the many-body language, we can proceed as follows (see \cite{Thermal},sect. 4.1 and \cite{KMS}, sect. 3).
\begin{bem} At the time of writing \cite{KMS} we were not aware of the related presentation in \cite{Thermal}. \cite{KMS} was primarily a follow-up paper of \cite{Collective}, in which some of the methods we shall use below have been developed.
\end{bem}

A system, placed in a heat bath, is already full of \tit{elementary excitations}. If we want to regard the equilibrium state as a Hilbert space vector, being invariant under the time evolution with energy normalized to zero, we can regard it as some sort of sea-level state, i.e., in the way of the \tit{Fermi-surface} being employed in solid-state physics. In \cite{Collective} we envoked the ingeneous \tit{Landau picture of deep-lying elementary excitations} which are supposed to interact only weakly.
\begin{conjecture} We assume that the many-particle system and its dynamics can be understood as being built up by weakly interacting long-lived collective or elementary excitations which are in general completely different from the elementary building blocks of the bulk system. This allows us to introduce a quasi-free Hamiltonian and creation- and annihilation operators of these new excitations (cf. \cite{Collective}).
\end{conjecture}
\begin{bem} In the older literature the corresponding transformation to new variables is frequently called a canonical transformation.
\end{bem}
\begin{ob} In a temperature state (which is already full of excitations) an ordinary (real-space) excitation mode of energy-momentum $(\omega(k),k)$ can be described in essentially two ways (in marked contrast to the ground-state). It can consist of an extra excitation above the Fermi-surface of positive $\omega(k)$ and momentum $k$ and/or the annihilation of a so-called hole below the Fermi-surface, the hole having energy-momentum $(-\omega(k),-k)$. In general an ordinary excitation (which we call in the following real-space) will be a temperature-dependent superposition of the two more primordial processes (cf. \cite{Thermal},sect. 4.1 and \cite{KMS}, sect. 3).
\end{ob}
\begin{bem} In the following we have to carefully distinguish between these two types of excitations. The real space excitations are in general observable while the more primordial excitations, making up the superposition are difficult to observe separately. It is perhaps even  impossible in the case of RQFT, which we shall discuss below.
\end{bem}

We now briefly recapitulate the discussion in \cite{Thermal} which is approximative in character by using the ordinary canonical QFT technique of canonical commutation relations and Fock-space machinery followed by canonical perturbation theory. In the next subsection we then describe the situation within the framework of Tomita-Takesaki and KMS-theory of general interacting QFT based on the paper \cite{KMS}. Using the notation of \cite{Thermal} we have for an ordinary real-space annihilation operator in a temperature representation with the following definition:
\begin{defi} We call the operators $a^{(+)}(k)$ real-space variables, the (temperature dependent) operators $a^{(+)}(k,\beta),\tilde{a}^{(+)}(k,\beta)$ primordial variables. 
\end{defi} 
\beq a(k)=\cosh c_k\cdot a(k,\beta)+\sinh c_k\cdot\tilde{a}^+(k,\beta)   \eeq
where it is assumed that $a(k),a(k,\beta),\tilde{a}(k,\beta)$ and their adjoints satisfy the same (bosonic) canonical commutation relations, e.g.:
\beq [a(k),a^+(l)]=\delta (k-l)   \eeq

$a(k,\beta)$ annihilates a mode above the Fermi surface, $\tilde{a}(k,\beta)$ annihilates a hole below the Fermi surface, that is, $\tilde{a}^+(k,\beta)$ creates a hole of energy-momentum $(-\omega_k,-k)$. This implies that the tilde operators commute with the $a^{(+)}(k,\beta)$.
\begin{bem} In the following $\omega_k$ is chosen positive.
\end{bem}
As the $a^{(+)}(k,\beta),\tilde{a}^{(+)}(k,\beta)$ operators are assumed to fulfill the canonical commutation relations separately, we can construct another representation in \tit{real space}, i.e., we define:
\beq \tilde{a}(k):=\cosh c_k\cdot \tilde{a}(k,\beta)+\sinh c_k\cdot a^+(k,\beta)   \eeq
\begin{ob} This tilde representation again commutes with the $a^{(+)}$-representation.
\end{ob}
\begin{bem} The coefficients $c_k$ are also functions of $\beta$ (see below). 
\end{bem}

The backtransform now reads:
\beq a(k,\beta)=\tilde{a}(k)\cdot\cosh c_k-\tilde{a}^+(k)\cdot\sinh c_k   \eeq
\beq \tilde{a}(k,\beta)=\tilde{a}(k)\cdot\cosh c_k+a^+(k)\cdot\sinh c_k  \eeq
\begin{bem} Similar relations hold for fermions.
\end{bem}
\begin{ob} In the primordial variables we have the following Hamilton operator:
\beq H_0=\int d^3k\, \omega(k,\beta)\cdot (a^+(k,\beta)a(k,\beta)-\tilde{a}^+(k,\beta)\tilde{a}(k,\beta))   \eeq
which reads in the real space variables:
\beq H_0=\int d^3k\, \omega(k,\beta)\cdot (a^+(k)a(k)-\tilde{a}^+(k)\tilde{a}(k))  \eeq
\end{ob}
\begin{bem} We see that via the finer analysis with the help of the primordial variables we get the kind of doubling we found in the Tomita formalism.
\end{bem}

We conclude this brief introduction into the \tit{thermo field framework} with giving the functional form of the $c_k$ (see \cite{Thermal}). 
\begin{postulate} The equilibrium state (a Hilbert space vector) $|0,\beta>$ is annihilated by $a(k,\beta),\tilde{a}(k,\beta)$:
\beq a(k,\beta)|0,\beta>=0=\tilde{a}(k,\beta)|0,\beta>   \eeq
It is hence not annihilated by the $a(k),\tilde{a}(k)$.
\end{postulate}
\begin{bem} Note the simlarity to the Unruh-blackhole-situation.
\end{bem}
\begin{ob} After some calculations (see \cite{Thermal}) we get:
\beq (\sinh c_k)^2=(e^{\omega(k)}-1)^{-1}=:f_B(\omega(k))   \eeq
That is, we have for example:
 \beq a(k)=(1+f_B^{1/2})\cdot a(k,\beta)+(f_B)^{1/2}\cdot\tilde{a}^+(k,\beta)  \eeq
 \beq\tilde{a}(k)=(1+f_B)^{1/2}\cdot \tilde{a}(k,\beta)+(f_B)^{1/2}\cdot a^+(k,\beta)  \eeq
\end{ob}
\subsection{The Return of the Dirac Sea -- The General Case}
In \cite{Cluster} we employed the natural two-sidedness of the energy-momentum spectrum of KMS-states (already suggesting a particle-hole picture). A more recent application is \cite{Lorentz}, where more references can be found. For a two-point function of the type $(\Omega|A(x,t)B\Omega)$ we have for the Fouriertransform $J(k,\omega)$ as a consequence of the KMS-condition:
\beq Re\, J(-k,-\omega)=e^{-\beta\omega}\cdot Re\, J(k,\omega)\; ,\;  Im\, J(-k,-\omega)=-e^{-\beta\omega}\cdot Im\, J(k,\omega)  \eeq

In the following we will use the results of \cite{KMS} to establish such a collective particle-hole picture within the regime of RQFT. An important technical tool is the spectral resolution of observables or general operators (a kind of Fourier-transform) with respect to the Tomita-KMS-evolution (for more details see \cite{KMS} and \cite{Collective}). We define $A(\omega)$ via
\beq \int A(s)\cdot f(s)\, ds=:\int A(\omega)\cdot\hat{f}(\omega)\, d\omega   \eeq
or formally (to be viewed in a distributional sense):
\beq A(\omega)=(2\pi)^{-1/2}\int e^{-is\omega}\cdot A(s)\, ds\; ,\; A(s)=(2\pi)^{-1/2}\int e^{is\omega}\cdot A(\omega)\, d\omega   \eeq
with
\beq (A^{\ast}(\omega)=(A(-\omega)^{\ast}   \eeq
which follows immediately from the preceding formulas. Physically it implies the following. If $\hat{f}(\omega)$ is, for example, concentrated around $\omega_0>0$ and thus as well the spectral support of  $\int A(s)\cdot f(s)\, ds$, the spectral support of $(\int A(s)\cdot f(s)\, ds)^{\ast}$ is peaked around $-\omega_0$. This means, physically it can be interpreted as a switch between particles and holes.
\begin{bem} We suppress the hat over $A$ in $A(\omega)$ because this notation would become quite cumbersome in the following formulas.
\end{bem}

The idea is now the following. Given an element $A\in\RO$ we have (due to the Reeh-Schlieder theorem) $A\Omega\neq 0\neq A^{\ast}\Omega$ (no local annihilation of the vacuum). We want now to construct a splitting of $A$ which resembles the preceding construction of creation/annihilation operators of collective excitation- or hole-modes. Using the above spectral decomposition of $A$, we want to have the following representation:
\begin{satz} There exist (distributional) operators $A^+(\omega),(A^+)^{\ast}(\omega)$ with
\beq A^+(\omega)\Omega=A(\omega)\Omega\; ,\;  (A^+)^{\ast}(\omega)\Omega=0   \eeq
For $\omega>0$ we associate  $A^+(\omega)$ with the creation of a collective excitation mode, $(A^+)^{\ast}(\omega)$ with its annihilation. For $\omega<0$ we regard them correspondingly as annihilation or creation operators of a hole in the Dirac sea, given by the KMS-state; i.e., in this case, the quantum vacuum with its vacuum fluctuations.
\end{satz}
\begin{bem} Note that we treat (in general) fully interacting models of RQFT. This implies that (as in the Landau picture) we cannot expect to have such a clean and simple situation as in the preceding subsection. As a consequence, the temperature dependent prefactors, we calculate in the following, are slightly different from the ones in \cite{Thermal} (a different normalization). Furthermore, starting from arbitrary observables $A$ we cannot expect to create creation/annihilation operators of single collective excitations by this method (note for example that we cannot employ a joint energy-momentum spectrum as we have extensively used in \cite{Collective}; we have only the Tomita evolution at our disposal). In general we might get clusters of such excitations. But, not that, in contrast to the preciding case, we analyze the patterns of vacuum fluctuations the behavior of which is supposed to be different from collective excitations in true many-body systems anyhow.
\end{bem}
For the proof we refer to \cite{KMS}.

We arrive at the following formulas:
\begin{ob} It holds
\beq A^+(\omega)=\left(-\frac{1}{1-e^{\omega}}\right)A(\omega)+\left(\frac{e^{\omega/2}}{1-e^{\omega}}\right)(A^{\ast})'(\omega)  \eeq
with $JAJ=:A'\in\RO'$ and correspondingly for $(A^+)^{\ast}(\omega)$. The backtransform reads:
\beq A(\omega)=\left(\frac{1-e^{-\omega}}{1+e^{-\omega}}\right)A^+(\omega)- \left(\frac{1-e^{-\omega}}{1+e^{-\omega}}\right)((A^{\ast})^+)^{\ast}(\omega)  \eeq
As in the preceding subsection we can represent observables lying in $\RO'$ in a related way as superpositions.
\end{ob} 
Proof: The representation for $A^+(\omega)$ can be found in \cite{KMS}. The formula for the backtransformation follows after some tedious but straightforward calculations.

Our task is now to analyze and explain the physics behind these (technical) results. To make the above formulas more transparent we simplify the notation slightly. 
\begin{ob} We have constructed four types of primordial operators:
i) $\omega>0$: $a^+(\omega,\beta),a^-(\omega,\beta)$, called modular (collective mode) creation/annihilation operators.\\
ii) $\omega<0$:  $b^+(\omega,\beta),b^-(\omega,\beta)$, called modular (collective mode) hole- creation/annihilation operators.\\
By reinterpreting the terms in the above formulas we see that, for example, $A(\omega)$ is a superposition of a modular mode creation and a hole annihilation operator.
\end{ob}
\begin{bem} We could and shall also call them modular particle/hole creation/annihilation operators whereas these are possibly not particles in the Minkowski sense. This is at least not clear at the moment (see below). In the Rindler case they are for example called Rindler particles. One should however note that in the Rindler or BH-scenario the reference state is the Rindler (or Boulware) Fock vacuum. Therefore we have (at first glance) no hole operator in that case.
\end{bem}

It is important to realize how these new operators are constructed. Each of them is a temperature dependent superposition of a term coming from $\RO$ and a term coming from $\RO'$, which is the causal complement of $\RO$. That is, in contrast to the box-KMS system, where all terms were located (at least at first glance) within the box, the corresponding terms in  the RQFT-scenario are geometrically separated. To understand the physical implications of this remarkable observation we want to refer to our above cited papers \cite{Wormhole},\cite{Quantum},\cite{Planck}, in which we analyzed the dynamics and long-range entanglement structure of the vacuum fluctuations.

In these papers we showed by an analysis which started from almost first and generally accepted principles that the vacuum fluctuations are strongly and long-range \tit{anticorrelated}, i.e. positive and negative fluctuations follow each other in a spatially very rigid pattern.
\begin{conclusion} In contrast to the excitations in ordinary Minkowski space like e.g. $A\Omega$ with $A\in\RO$, which are essentially localized  in some finite region of S-T, the primordial modular excitations extend over the whole Minkowski space with, for example, 
\beq \int A^+(\omega)\cdot \hat{f}(\omega)\, d\omega =\int A^+(s)\cdot f(s)\, ds  \eeq
being the superposition of a contribution coming from $\RO$ and and a term coming from $\RO'$. On the other hand, the essentially localized $A\in\RO$ is the superposition of two translocal contributions (as can be seen from the above formulas). These two terms do appropriately interfere with each other to generate a Minkowski-localized observable.
\end{conclusion}
\section{Conclusion--The Coherent Underlying Picture}
We want to put what we have learned so far into a coherent picture. On the one hand we have the universal but context dependent thermal (KMS) behavior of the relativistic quantum vacuum, if reduced to a finite region of S-T. As a consequence of some rigorous results of RQFT (the vacuum as a cyclic and separating vector) we can establish a link to the Tomita-Takesaki-theory of v.Neumann algebras. This latter theory has as key ingredients the so-called Tomita-involution $J$, which relates the algebras $\RO$ to its commutant $\RO'$ and the Tomita- (or modular) operator $\Delta$, with the corresponding evolution $\Delta^{is}\circ\Delta^{-is}$ leaving invariant the algebras $\RO$ and $\RO'$. The preceding sections show how this rigid and universal structure leads (almost inevitably) to a somewhat hidden but more primordial structural level below the \tit{causal} and \tit{local} surface structure being based on the Minkowski space. This second level is the place where the vacuum fluctuations and their dynamics hold sway. 

Our analysis led to a picture where the patterns of these (basically non-locally behaving) vacuum fluctuation induce, via appropriate superpositions, the local behavior in Minkowski space. With the help of some reasonable assumptions we undertook to characterize this underlying structure, employing the picture of collective or elementary excitations which has been so successful in many-body physics (cf. \cite{Collective}). Note that we developed such ideas already in \cite{Quantum}; such ideas were also developed in \cite{Bohm} p.92 ff. That is, we argue that these patterns of vacuum fluctuations can be described (at least in an approximative way) as being generated by elementary modular particle- (better: mode-) and hole-excitations within a medium (the quantum vacuum) which contains already a distribution of these excitations. If one now reverses the steps of our analysis one can conclude the following:    
\begin{conclusion} The Tomita-KMS-structure on the Minkowski-space level derives from the above described underlying more primordial structure.\\
i) As described and analyzed in sect. 5.1 and 5.2, the duality symmetry between $\RO$ and $\RO'$, mediated by the antiunitary operator $J$, is a consequence of the fact that in the underlying structure we have more DoF (degrees of freedom), that is, particle-hole excitations , so that via appropriate superpositions we can compose from the non-local modular particle-hole creation/annihilation operators not only the localized objects lying in $\RO$ but, with the help of another conjugate superposition, also the corresponding objects lying in $\RO'$.\\
ii) The Tomita evolution  $\Delta^{is}\circ\Delta^{-is}$, leaving $\RO$ and $\RO'$ invariant, can be understood as the respective reductions of the corresponding non-local evolution of the patterns of vacuum fluctuations.\\
iii) The whole structure carries the flavor of the old Dirac-Sea picture.
\end{conclusion}
\begin{bem} The following is perhaps an interesting observation in this context. In \cite{Collective} we heavily exploited the existence of space-time symmetry which implied the existence of a joint energy-momentum spectrum. By this means we were able to distill dispersion laws of long-lived collective excitation branches out of the full energy-momentum spectrum. In the case of the vacuum fluctuations we have only the one-dimensonal Tomita evolution at our disposal. As a consequence the corresponding conjectured collective excitations have no particular energy- momentum dependence (i.e., a dispersion law). Quite to the contrary they may more resemble standing waves which do not propagate through space. This is perhaps an interesting aspect concerning the question of the nature of Rindler particles.
\end{bem}

This structure becomes particularly transparent if we employ the operator algebras $\M$ and $\M'$. We learned that we can generate these particular algebras by adding appropriate boundary operators to the algebra $\RO$ (cf. sect.4). We then get a quite familiar thermodynamical structure as we know it from ordinary quasi-isolated systems. Most notably, we make visible on the level of Minkowski space the fine structure of the vacuum fluctuation patterns in form of the so-called `pieces of the vacuum'. In this particular context we want to address as a last point the \tit{thermal-time conjecture} of Connes and Rovelli. The example of a non-equilibrium situation discussed in the introduction and the emergence of boundary conditions in the construction of $\M$ suggest the following point of view.
\begin{ob}[The Thermal-Time Hypothesis] 
While at first glance the universal existence of a (modular) automorphism group for each algebra of observables $\RO$ may be regarded as a surprising but perhaps abstract result without a deeper physical content in most cases, the preceding remarks may lead to the following idea. We suggest to consider the evolution parameter `$s$' and the Tomita Hamiltonian $K$ as a \tit{thermal time} and a local Hamiltonian, havng their origin in the dynamics of the vacuum fluctuations, while time and Hamiltonian on the level of Minkowski space, i.e. in RQFT, are rather of the nature of \tit{effective Hamiltonians} and a corresponding effective time concept. The addition of boundary conditions create localized eigenvectors and a localized time evolution in contrast to the global Minkowski time evolution (based on a mechanical/macroscopic time concept). 
\end{ob}


\begin{thebibliography}{99}     
  {\small
\bibitem{1} W.G.Unruh: PR D 14(1976)870
\bibitem{2}  N.D.Birell,P.C.Davies: ``Quantum fields in curved space'', Cambridge Univ.Pr., Cambridge 1982
\bibitem{3} D.W.Sciama,P.Canelas,D.Deutsch: Adv.Phys. 30(1981)327
\bibitem{4} W.G.Unruh,R.M.Wald: PR D 29(1984)1047
\bibitem{5} G.L.Sewell: Ann.Phys. 141(1982)201 
\bibitem{6} D.Buchholz,C.Solveen: ``Unruh Effect and the Concept of Temperature'', arXiv:1212.2409
\bibitem{7} M.Requardt: ``The Concept of Unruh Temperature or What is real. A Comment to a paper by Buchholz and Solveen''
\bibitem{8} J.J.Bisognano,E.H.Wichmann: J.Math.Phys. 16(1975)985, J.Math.Phys. 17(1976)303
\bibitem{9} R.Haag: ``Local Quantum Physics'', Springer, Berlin 1996
\bibitem{10} R.F.Streater,A.S.Wightman: ``PCT, Spin and Statistics, \& All That'', Benjamin, N.Y. 1964
\bibitem{11} M.Takesaki. ``Tomita's Theory of Modular Hilbert Algebras'', LNM 128, Springer, Berlin 1970
\bibitem{12} O.Bratteli,D.W.Robinson: ``Operator Alge P.Martinetti,bras and Quantum Statistical Mechanics I, Springer, Berlin 1979 
\bibitem{13} A.Connes,C.Rovelli: Class.Quant.Grav. 11(1994)2899
\bibitem{14} C.Rovelli: Class.Quant.Grav. 10(1993)1549
\bibitem{Martinetti} P.Martinetti,C.Rovelli: Class.Quant.Grav. 20(2003)4919
\bibitem{15} M.Requardt: ``QFT in restricted Domains of Minkowski Space. The Thermalization of the Physical Vacuum as the Physics of certain Open Systems'', Goettingen preprint, unpublished 1991
\bibitem{16} T-.T.Paetz: ``An Analysis of the Thermal-Time Concept of Connes and Rovelli'', diploma thesis, Goettingen 2010, http://www.theorie.physik.uni-goettingen.de//forschung/qft/theses/dipl/Paetz.pdf
\bibitem{17} D.W.Robinson: Comm.Math.Phys. 31(1973)171
\bibitem{18} R.V.Kadison,J.R.Ringrose: ``Fundamentals of the Theory of Operator Algebras'' Vol.II, Academic Pr. Orlando 1986
\bibitem{Wald} R.M.Wald: ``Quantum Field Theory in Curved Space-Time'', The University of Chicago Pr., Chicagoß 1994
\bibitem{Kay1} B.Kay: Comm.math.Phys. 100(1985)57
\bibitem{Summers} S.J.Summers: Rev.Math.Phys. 2(1990)201
\bibitem{Dixmier} J.Dixmier: ``Les Algebres D'Operateurs Dans L'Espace Hilbertien'', Gauthier-Villars, Paris 1957
\bibitem{Jones} V.J.R.Jones: ``Von Neumann Algebras'', Lecture Notes 2009, see Homepage of V.Jones or math.berkeley.edu/~vfr(2009) 
\bibitem{Longo} S.Doplicher,R.Longo: Invent.math. 75(1984)493
\bibitem{Neumann} J.v.Neumann: ``Mathematische Grundlagen der Quantenmechanik'', Springer, Berlin 1968
\bibitem{Requ} M.Requardt: ``Higher-Order-Schmidt-Representations and their Relevance for the Basis Ambiguity'', arXiv:1009.1223
\bibitem{Bratteli} O.Bratteli,D.W.Robinson: ``Operator Algebras and Quantum Statistical Mechanics II, Springer, Berlin 1981 
\bibitem{Wheeler} C.W.Misner,K.S.Thorne,J.A.Wheeler: ``Gravitation'', Freeman, San Francisco 1970
\bibitem{Kay2} B.Kay: Helv.Phys.Acta 58(1985)1030
\bibitem{Wormhole} M.Requardt: ``Wormhole Spaces: the Common Cause rfor the Black Hole Entropy-Area Law, the Holographic Principle and Quantum Entanglement'', arXiv:0910.4017
\bibitem{Quantum} M.Requardt: ``Quantum Theory as emergent from an undulatory translocal Sub-Quantum Level'', arXiv:1205.1619
\bibitem{Planck} M.Requardt: Mod.Phys.Lett. A 22(2007)791, arXiv:gr-qc/0505019
\bibitem{Ado1} M.Redhead: Found.Phys. 25(1995)123
\bibitem{Ado2} S.J.Summers: ``Yet more Ado about Nothing: The Remarkable Relativistic Vacuum State'', arXiv:0802.1854 
\bibitem{Thermal} H.Umezawa,H.Matsumoto,M.Tachiki: ``Thermo Field Dynamics'', North-Holland, Amsterdam 1982
\bibitem{Schroer} B.Schroer:  ``More on area density of localization entropy'', arXiv:hep-th/0511291
\bibitem{Araki1} H.Araki,E.J.Woods: J.Math.Phys. 4(1963)637
\bibitem{Araki2} H.Araki,W.Wyss: Helv.Phys.Acta 37(1964)136
\bibitem{Ojima} I.Ojima: Ann.Phys. 137(1981)1
\bibitem{KMS} M.Requardt: J.Phys. A:Math.Gen. 18(1985)287
\bibitem{Collective} H.Narnhofer,M.Requardt,W.Thirring: Comm.Math.Phys. 92(1983)247
\bibitem{Cluster} M.Requardt: J.Phys. A:Math.Gen. 13(1980)1769
\bibitem{Lorentz} M.Requardt: ``Spontaneous Symmetry Breaking of Lorentz and Galilei Boosts in Relativistic Many-Body Systems'', arXiv:0805.3022
\bibitem{Bohm} D.Bohm: ``Wholeness and the Implicate Order'', Routledge and Kegan, London 1980

}
\end{thebibliography}
\end{document}